\documentclass[twocolumn,amsmath,amssymb]{revtex4-1}

\usepackage{epsf}
\usepackage{amsmath}
\usepackage{amssymb}
\usepackage{amsbsy}
\usepackage{verbatim}
\usepackage{graphicx}
\usepackage{datetime}
\usepackage{subfig}
\usepackage{mathrsfs} 
\usepackage[retainorgcmds]{IEEEtrantools} 
\usepackage[utf8x]{inputenc}
\usepackage{slashbox}

\usepackage[colorlinks=true, linkcolor=blue, citecolor=blue, filecolor=blue, runcolor=blue, urlcolor = blue]{hyperref}
\usepackage[usenames,dvipsnames]{color}

\newcommand{\mb}[1]{{\mathbf{#1}}}

\definecolor{issuePJA_color}{rgb}{1.0,0.0,0.0}

\definecolor{commentPJA_color}{rgb}{1.0,0.0,0.8}

\DeclareGraphicsExtensions{.png}
\begin{document}

\preprint{Dynamic Implicit Solvent Coarse-Grained Models of Lipid Bilayer Membranes : Fluctuating Hydrodynamics Thermostat} 

\title{Dynamic Implicit-Solvent Coarse-Grained Models of Lipid Bilayer Membranes : Fluctuating Hydrodynamics Thermostat}

\author{Yaohong Wang}
\affiliation{Department of Mathematics \\
University of California Santa Barbara.}

\author{Jon Karl Sigurdsson}
\affiliation{Department of Mathematics \\
University of California Santa Barbara.}

\author{Paul J. Atzberger}
\thanks{\textit{address:} University of California;
Department of Mathematics; Santa Barbara, CA 93106;
\textit{e-mail:} atzberg@math.ucsb.edu; 
\textit{web:} \href{http://www.atzberger.org/}{www.atzberger.org};
\textit{phone:} 805 - 679 - 1330.}
\affiliation{Department of Mathematics, \\
University of California Santa Barbara.
\\
\\
}

\begin{abstract}
\noindent
\textbf{Abstract:}
\noindent
Many coarse-grained models have been developed for equilibrium studies of 
lipid bilayer membranes.  To achieve in simulations access to length-scales and time-scales 
difficult to attain in fully atomistic molecular dynamics, these coarse-grained 
models provide a reduced description of the molecular degrees of freedom and often 
remove entirely representation of the solvent degrees of freedom.  In such implicit-solvent 
models the solvent contributions are treated through
effective interaction terms within an effective potential for the free energy.  For
investigations of kinetics, Langevin dynamics is often used.  
However, for many dynamical processes within bilayers this approach is insufficient 
since it neglects important correlations and dynamical contributions that are missing 
as a result of the momentum transfer that would have occurred through the solvent.  To 
address this issue, we introduce a new thermostat based on fluctuating hydrodynamics 
for dynamic simulations 
of implicit-solvent coarse-grained models.  Our approach couples the coarse-grained 
degrees of freedom to a stochastic continuum field that accounts for both the solvent 
hydrodynamics and thermal fluctuations.  We show our approach captures important  
correlations in the dynamics of lipid bilayers that are missing in simulations performed using  
conventional Langevin dynamics.  For both planar bilayer sheets and bilayer vesicles, we investigate 
the diffusivity of lipids, spatial correlations, and lipid flow within the bilayer.  The 
presented fluctuating hydrodynamics approach provides a promising way to extend implicit-solvent 
coarse-grained lipid models for use in studies of dynamical processes within bilayers.
\end{abstract}

\maketitle



\section{Introduction}
\label{Sec_Introduction}

Many coarse-grained models have been developed for equilibrium studies of 
lipid bilayer membranes~\citep{Cooke2005,Cooke2005a,Wang2010,Brannigan2006,Farago2003,Goetz1998a,Revalee2008,Wang2005a,Drouffe1991,Bourov2005, B.Smit1993, Stevens2004, Noguchi2001}.  To achieve in simulations access to length-scales and time-scales 
difficult to attain in fully atomistic molecular dynamics, these coarse-grained lipid
models have been developed to provide a reduced description of the molecular degrees 
of freedom and often remove entirely representation of the solvent degrees of 
freedom~\citep{Cooke2005,Cooke2005a,Wang2010,Farago2003,Brannigan2006,Revalee2008,Noguchi2001}.
In such implicit-solvent models, the solvent contributions are treated through
effective interaction terms within an effective potential for the free energy.
Dynamical processes are then often investigated using  
Langevin dynamics~\citep{Revalee2008,Cooke2005,Cooke2005a,Noguchi2001}.
However, for many problems this approach is insufficient since it neglects
important correlations and dynamic contributions from the momentum transfer 
that would have occurred through the solvent.  To incorporate these effects,
we introduce a thermostat based on fluctuating hydrodynamics for dynamic 
simulations of implicit-solvent coarse-grained models.  Our approach couples the 
coarse-grained degrees of freedom to a stochastic continuum field that accounts for 
both the solvent hydrodynamics and thermal fluctuations.   We present a general mathematical 
framework and specific methods for how to couple these descriptions in a manner 
consistent with statistical mechanics and in a manner amenable to efficient computational 
methods~\citep{AtzbergerSELM2011, Atzberger2007a}.  
We then present a number of results for dynamical properties of the fluctuating 
hydrodynamics bilayer model.  We present results for both self-assembled planar bilayers 
and self-assembled vesicles.  In particular, we consider the relaxation of the mean-squared displacement characterizing diffusivity of 
lipids within the bilayer for both the fluctuating hydrodynamics method and conventional Langevin dynamics.  
We next consider the correlations between the motions of an individual lipid and those of its neighbors 
within a patch of varying size.  Finally, we consider the pair correlations for the motions of lipids within the 
bilayer.  We find interesting vortex-like flow structures for the correlated lipid motions within the bilayer
that are similar to those observed in explicit solvent bilayer simulations.
We expect the introduced fluctuating hydrodynamics methods to provide powerful new approaches for performing dynamical 
studies utilizing implicit-solvent coarse-grained models.

\section{Implicit-Solvent Coarse-Grained Lipid Model}
\label{Sec_ImplicitSolvent}

Many coarse-grained implicit-solvent models have been developed for equilibrium
studies of lipid bilayer membranes~\citep{Cooke2005,Cooke2005a,Wang2010,Farago2003,Goetz1998a,Revalee2008,Wang2005a,Drouffe1991}.  
These models capture at different levels of resolution the molecular details of lipids.  To demonstrate our 
approach, we shall focus on the specific coarse-grained lipid model developed by Cooke 
and Deserno~\citep{Cooke2005,Cooke2005a}.  
\begin{figure}[h]
\centering
  \includegraphics[width=8.5cm]{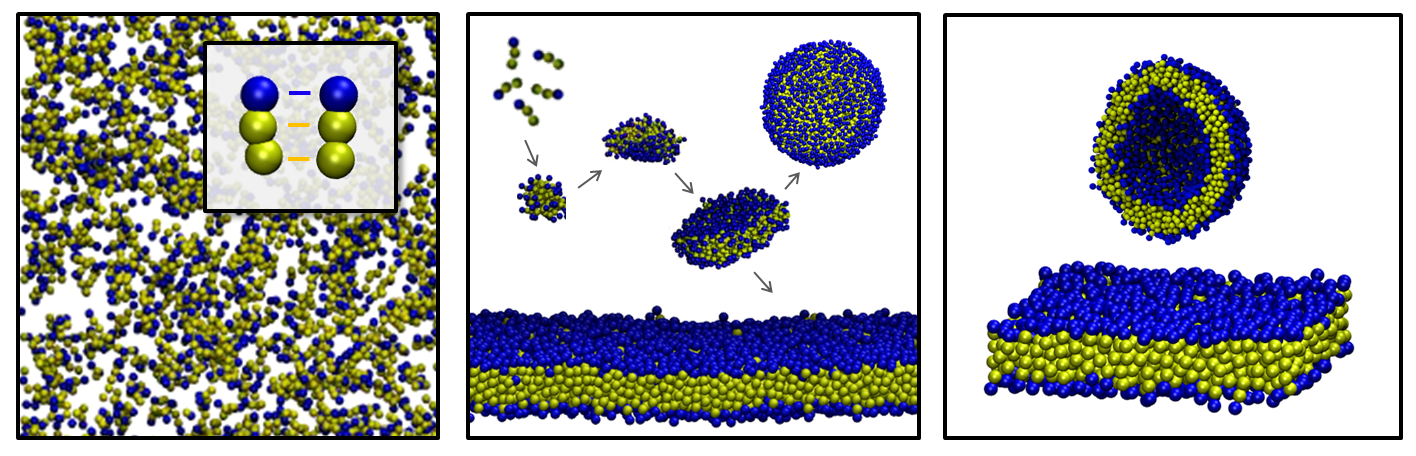}
\caption{Implicit-Solvent Coarse-Grained Lipid Model : Individual lipids are modeled by  
three coarse-grained units that interact to mimic amphiphilic molecules~\citep{Cooke2005a}.
Starting from a concentrated solution of model lipids, structures are self-assembled (left).
The self-assembly proceeds initially through the formation of aggregates that resemble micelles 
which subsequently merge into larger disk-like structures (middle).  These structures merge further and 
after reaching a critical size they either orient to span the entire periodic domain as a planar bilayer 
or wrap spontaneously to form a vesicle (middle and right).  
}
\label{fig_CG}  
\end{figure}
In this model each lipid is 
represented by three coarse-grained units.  The first unit accounts 
for the polar hydrophilic head group of the lipid and the remaining two
units account for the hydrophobic groups along the hydrocarbon tail 
of the lipid, see Figure~\ref{fig_CG}.  An effective interaction potential 
is developed that 
takes into account excluded volume interactions, van der Waal's attraction,
and the hydrophobic-hydrophilic effect.  The solvent mediated 
effects that drive formation of the bilayer 
structure is taken into account through a long-range attractive 
interaction term between the tail units in the coarse-grained model.
To obtain robust bilayers that exhibit a fluid phase, it was found 
important to use an attractive interaction that has a broad energy 
well~\citep{Cooke2005a}.

Extensive studies have been performed to parameterize this lipid model 
to obtain reasonable equilibrium properties, such 
as the bilayer bending elasticity, compression modulus, average area per lipid, 
and tension~\citep{Cooke2005a}.  
Throughout, we shall use the specific parameterization given in Table~\ref{table_params}.
An important feature of the lipid model is the self-assembly of stable bilayer structures
from a solution of lipids.  For our studies, we have performed simulations over long trajectories 
to self-assemble both planar bilayer sheets and vesicles, see Figure~\ref{fig_CG}.  For the 
planar bilayer case with periodic boundary conditions, special care must be taken to 
obtain an appropriate domain size to ensure a small surface tension of the constructed 
bilayer.  We have implemented a modified Andersen thermostat/barostat to equilibrate our planar bilayers 
in the $\gamma N V T$ ensemble with zero surface tension $\gamma = 0$, see~\citep{Andersen1980}.  We use in our dynamic studies 
a fixed domain having the average of the sampled domain sizes.  In the case of the vesicle bilayer,
no such special thermostatting was required.  On a large domain, we simulated a concentrated 
solution of lipids over several distinct long trajectories.  In a subset of these trials  
we obtained spontaneous self-assembly of vesicles, see Figure~\ref{fig_CG}.

\begin{table}
\begin{tabular}{|l|l|l|}
\hline
\textbf{Parameter}  & \textbf{Description}  & \textbf{Value} \\
\hline
$\sigma$   & lipid radius            & 1.0 nm \\
$\epsilon$ & energy scale            & 2.5 kJ$\cdot$M$^{-1}$ \\
$m_0$      & reference mass          & 1 amu \\
$w_c$      & energy potential width  & $1.2 \sigma$ \\
$m$        & lipid mass              & 720 $m_0$ \\
$\tau$     & time-scale              & $\sigma\sqrt{m_0/\epsilon}$ = 0.6 ps\\
$k_B{T}$   & thermal energy          & 1.0 $\epsilon$\\
$\rho$     & solvent mass density    & 602 $m_0/\sigma^3$ \\
$\mu$      & solvent viscosity       & 383 $m_0/\tau\sigma$ \\ 
$\Upsilon$ & drag coefficient        & 7210 $m_0/\tau$  \\
\hline
\end{tabular}
\caption{Parameterization of the Fluctuating Hydrodynamics Lipid Model}
\label{table_params}
\end{table}

\section{Fluctuating Hydrodynamics Thermostat}
\label{Sec_FluctHydro}

To account for the contributions of momentum transfer through the missing solvent 
degrees of freedom, we introduce a continuum stochastic field for the solvent that 
accounts for both hydrodynamics and thermal fluctuations.  For this purpose,
we introduce the following fluctuating hydrodynamic equations
\begin{eqnarray}
\label{equ_fld}
\rho\frac{d\mb{u}}{dt} & = & \mu \Delta{\mb{u}} -\nabla{p} 
+ \Lambda\left\lbrack \Upsilon \left(\mb{V} - \Gamma{\mb{u}}\right)\right\rbrack 
+ \mb{f}_{thm} \\
\label{equ_incomp}
\nabla \cdot \mb{u} & = & 0 \\
\label{equ_prt_vel}
m\frac{d\mb{V}}{dt} & = & -\Upsilon \left(\mb{V} - \Gamma{\mb{u}}\right) -\nabla \Phi + \mb{F}_{thm} \\
\label{equ_prt_pos}
\frac{d \mb{X}}{dt} & = & \mb{V}.
\end{eqnarray}
The $\mb{u}$ denotes the velocity of the 
solvent fluid and $\mb{X}, \mb{V}$ denotes the collective positions and velocities of the coarse-grained particles.
The $p$ denotes the pressure that imposes the incompressibility of the solvent fluid, $\nabla \cdot \mb{u} = 0$.
The $\rho$ is the mass density of the solvent, $\mu$ is the shear viscosity, $m$ is the mass matrix of the coarse-grained 
particles, and $\Phi$ is the potential energy of the coarse-grained particles.  
The thermal fluctuations are taken into account 
through the stochastic driving fields $\mb{f}_{thm}$
and $\mb{F}_{thm}$.  These are $\delta$-correlated Gaussian random fields  
with covariances
\begin{eqnarray}
\left\langle \mb{f}_{thm}(s) \mb{f}_{thm}(t)^T \right\rangle & = & -2k_B{T}\left(\mu\Delta - \Lambda\Upsilon\Gamma\right)\hspace{0.03cm}\delta(t - s)
\hspace{0.4cm} \\
\left\langle \mb{F}_{thm}(s) \mb{F}_{thm}(t)^T \right\rangle & = & 2k_B{T}\Upsilon\hspace{0.06cm}\delta(t - s) \\
\label{equ_fF_corr}
\left\langle \mb{f}_{thm}(s) \mb{F}_{thm}(t)^T \right\rangle & = & -2k_B{T}\Lambda\Upsilon\hspace{0.03cm}\delta(t - s).
\end{eqnarray}
Through the operators $\Upsilon$ and $\Gamma$, the term $-\Upsilon \left(\mb{V} - \Gamma{\mb{u}}\right)$ provides an 
effective coarse-grained model of how the local fluid flow exerts a drag force on the coarse-grained particles.  
\begin{figure}[h]
\centering
  \includegraphics[width=8cm]{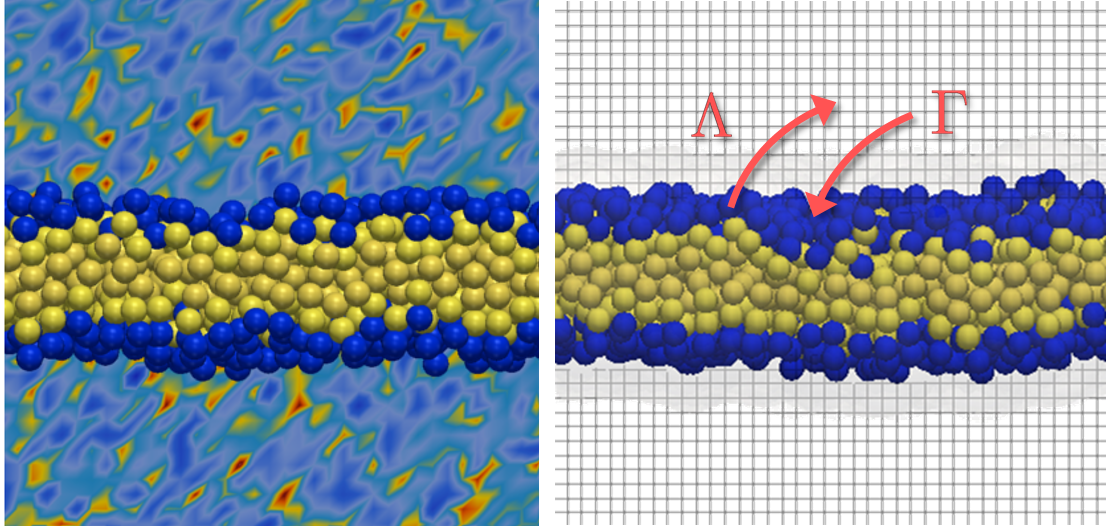}
\caption{Fluctuating Hydrodynamics Thermostat.  The stochastic velocity field of 
the solvent fluid during the bilayer simulation (left).  The Eulerian fluctuating hydrodynamic 
fields of the 
solvent are coupled to the Lagrangian degrees of freedom of the lipids
through the operators $\Lambda$ and $\Gamma$ (right).
}
\label{fig_CG}  
\end{figure}
The $\Lambda$ operator then models for the drag force how the equal-and-opposite forces exerted on the solvent are 
spatially distributed within the fluid body.  An important condition is that the coupling operators be adjoints 
$\Gamma = \Lambda^*$, see~\citep{AtzbergerSELM2011}.  This adjoint condition ensures the dissipation occurs only through the $\Upsilon$ drag 
and not as a consequence of the interconversion operators $\Gamma, \Lambda$~\citep{AtzbergerSELM2011}.  For 
the thermal fluctuations, the adjoint condition also greatly simplifies the form of the correlations 
of the stochastic driving fields and the algorithms needed for their computational 
generation~\citep{AtzbergerSELM2011}.  We refer to the coarse-grained fluctuating hydrodynamics approach 
of equations~\ref{equ_fld}--~\ref{equ_fF_corr} and related computational methods as the 
Stochastic Eulerian Lagrangian Method (SELM).  The key feature of this approach is the mixed use of 
an off-lattice Lagrangian description of the lipids that is coupled to an on-lattice Eulerian 
description of the solvent fluid.  This feature allows for leveraging numerical approaches 
from computational fluid dynamics and from molecular dynamics to perform efficient 
simulations of the coarse-grained fluctuating hydrodynamics model.  

Throughout, we shall use the specific coupling operators 
\begin{eqnarray}
\Gamma \mb{u}  & = & \int_{\Omega} \eta\left(\mb{y} - \mb{X}(t)\right) \mb{u}(\mb{y},t) d\mb{y} \\
\Lambda \mb{F} & = & \eta\left(\mb{x} - \mb{X}(t)\right)\mb{F}.
\end{eqnarray}
While other choices are possible and may be desirable, this approach is based on the 
Stochastic Immersed Boundary Method~\cite{Atzberger2007a}.  We use the kernel functions 
$\eta(\mb{z})$ chosen to be the Peskin $\delta$-Function given in~\citep{Peskin2002}.  This 
choice is made instead of the Dirac $\delta$-Function to ensure a model in which the  
mobility of individual particles have a finite effective hydrodynamic radius within the 
fluid~\cite{Atzberger2007a}.  This choice also has important numerical properties that ensure to a good
approximation translational invariance of the coupling despite the breaking of this
symmetry by the discretization lattice of the fluid~\citep{Atzberger2007a, Peskin2002}.  
While other choices of the coupling operators are possible, the Stochastic 
Immersed Boundary Method has been shown to provide a computationally efficient method for 
obtaining correct far-field hydrodynamic correlations and has a well-characterized  
near-field interaction~\citep{Atzberger2006b,Atzberger2007a,Atzberger2007c,Atzberger2009}.

\section{Comparison with Langevin Dynamics}
\label{Sec_Dynamic_Properties}

The SELM fluctuating hydrodynamics thermostat is compared to the
conventional Langevin dynamics by considering two cases.  In
the Langevin dynamics, we consider the case when the drag coefficient
is comparable to the Stokes drag of a particle of size $\sigma$ immersed 
in water, $\Upsilon = 7210$ $m_0/\tau$.  We also consider the case 
corresponding to many Langevin simulations in the literature 
where the drag coefficient is taken artificially small  
to achieve efficient equilibration and sampling, 
$\Upsilon = 0.06$ $m_0/\tau$~\citep{Cooke2005,Cooke2005a}.
For the reference Lennard-Jones units $\sigma, \epsilon, m_0$, 
see Table~\ref{table_params}.  
We remark that for the small drag case the dynamics are in an 
inertial regime exhibiting a coherent velocity on the time-scale 
for collisions to occur between the lipids.  In this regime, the collisions 
are expected to result in significant momentum transfer between 
the lipids before the momentum is suppressed by the Langevin drag.  
These two cases provide a useful baseline for comparison to 
the momentum conserving SELM thermostat.  
Throughout the SELM simulations, we use the solvent-lipid 
coupling strength determined by the Stokes drag 
in water, $\Upsilon = 7210$ $m_0/\tau$.

\subsection{Lipid Diffusivity within Planar Bilayers and Vesicles}

\begin{figure}[h]
\centering
  \includegraphics[height=6cm]{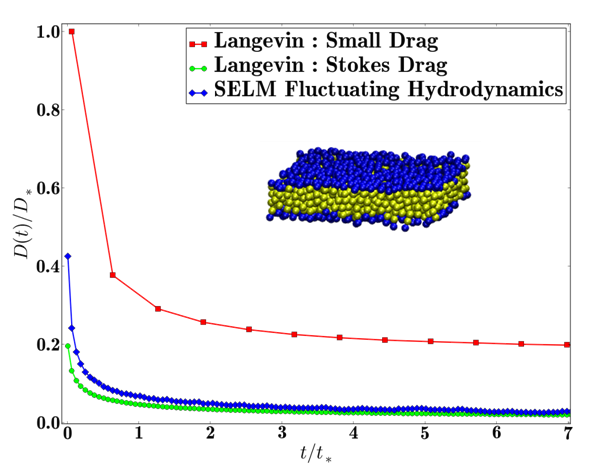}
  \caption{Lipid Diffusivity within a Planar Bilayer.}
\label{fig_diffusivity_planar}
\end{figure}

\begin{figure}[h]
\centering
  \includegraphics[height=6cm]{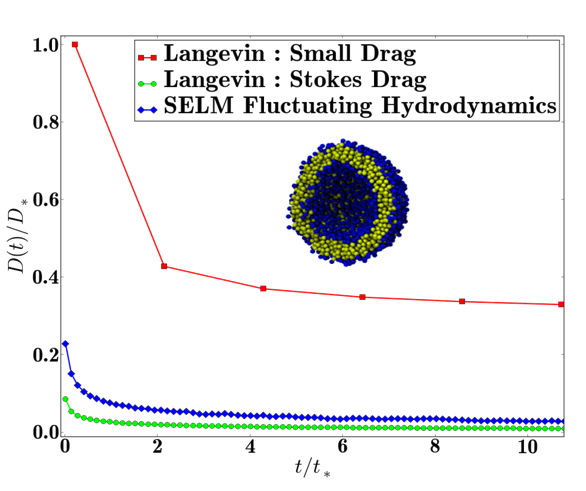}
  \caption{Lipid Diffusivity within a Vesicle Bilayer.}
\label{fig_diffusivity_vesicle}
\end{figure}

We initially study the diffusivity of lipids within the bilayer
using the conventional Langevin dynamics and then make a comparison
with the SELM dynamics.  The diffusivity considered is $D(t) = \langle\mathbf{X}^2(t)\rangle/t$
and non-dimensionalized by the small drag Langevin diffusivity $D_*(0)$
and time-scale $t_* = \sigma^2/D_*(0)$.
We perform studies of the diffusivity for 
both a planar bilayer and a vesicle.
We find in both cases 
that the SELM fluctuating hydrodynamics thermostat 
exhibits marked differences with the Langevin dynamics, 
see Figure~\ref{fig_diffusivity_planar} and~\ref{fig_diffusivity_vesicle}.
In the case of the Langevin dynamics with Stokes drag, we 
see that while the SELM dynamics yields a comperable diffusivity,
the relaxation to the steady-state occurs over a significantly longer 
time-scale.  This is 
consistent with the broad spectrum of time-scales associated with the 
relaxation of the individual hydrodynamic modes.  
For the fluctuating hydrodynamics thermostat, the well-known $t^{-3/2}$ decay 
of the velocity autocorrelation function can be established analytically in some 
limiting regimes and has been demonstrated in simulations~\citep{Atzberger2006b}.  
As expected, compared to the Langevin dynamics with small drag, there is a significant 
reduction in the diffusivity under the SELM dynamics.  However, compared to the 
Langevin dynamics with the Stokes drag coefficient, we find that the SELM dynamics
yields lipids with a somewhat larger diffusivity.  In this regime, we expect 
the diffusivity to be a consequence of the dissipation associated with the 
collective internal rearrangement modes of the lipids within the bilayer.  
We think the larger diffusivity under the SELM dynamics can be explained by 
two related mechanisms.  The first is the local conservation of momentum that 
transmits momentum between lipids inducing local correlations and flows.  The 
second is that the drag force acting on lipids depends only on the relative 
difference in velocity between the hydrodynamic field and the lipids.
This allows for rearrangements that are less dissipative when there is a 
coordinated deformation in the modes of the bilayer and fluid body.  This is 
in contrast to the Langevin dynamics where momentum is ceded to an undeformable fixed 
ambient medium.  Interestingly, as we shall discuss in latter sections, this result 
is somewhat in opposition to the alternative intuition that the obtained diffusivity 
under SELM dynamics might be smaller than the Langevin Stokes case as a consequence of 
the hydrodynamics giving additional local correlations.  The hydrodynamics is expected 
to result in coupling of nearby lipids
giving a type of coherent motion over a patch of the bilayer~\citep{Apajalahti2010}.

\subsection{Correlations between Lipids within a Bilayer Patch}

\begin{figure}[h]
\centering
  \includegraphics[width=9cm]{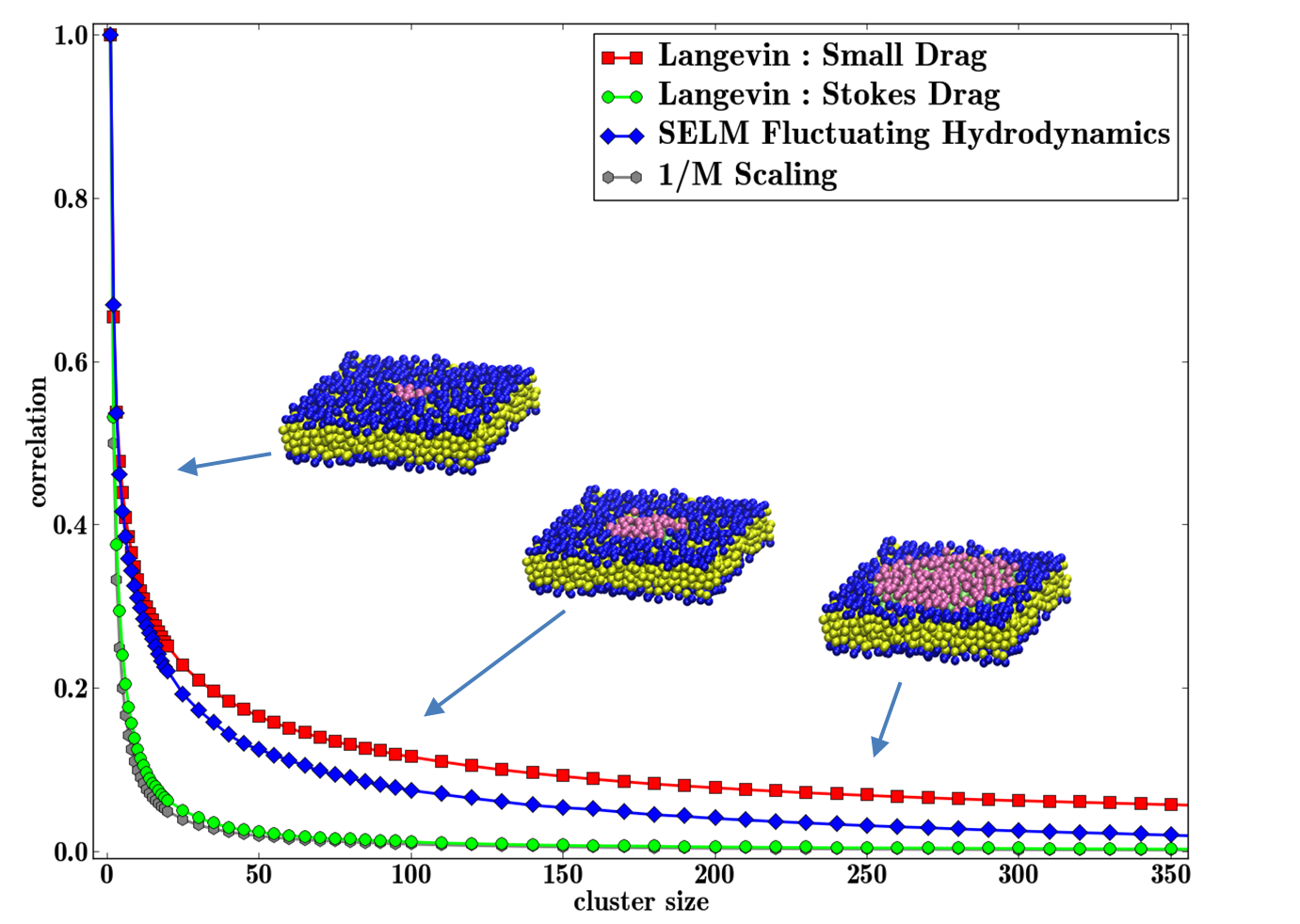}
  \caption{Correlations between Lipids within a Bilayer Patch}  
\label{fig_cluster}
\end{figure}

To characterize the correlations between lipids and those of its surrounding
neighborhood, we consider the motion of an individual lipid and a patch.
In particular, we consider for the displacement $\Delta_0{X}$
of a given reference lipid over a time $\delta{t}$ and its correlation to the 
displacement $\Delta_M{X}$ of the center-of-mass of a patch consisting of 
the $M$ nearest neighbors.  We consider the specific correlation 
$\left \langle \Delta_0{X} \Delta_M{X} \right \rangle/\left \langle \Delta_0{X}^2 \right \rangle$.
The results for the SELM fluctuating hydrodynamics and Langevin dynamics 
are shown in Figure~\ref{fig_cluster}.  
For lipids that diffuse with only short-ranged correlations 
with their neighbors we expect a base-line correlation as 
the patch size grows to scale like $\sim 1/M$.  The Langevin 
dynamics with small drag exhibits significant long-range 
correlations.  In this case, the lipid dynamics exhibit 
significant inertial effects and momentum is transferred through 
collisions.  This results in correlations between nearby lipids 
that are longer-ranged.  In contrast, the Langevin dynamics having the Stokes 
drag coefficient exhibits the $1/M$ scaling to a good approximation.  This 
indicates that correlations in the displacements of the lipids within
the bilayer are strongly suppressed by the Langevin drag.  For the SELM dynamics 
using the same Stokes drag coefficient, we find long-range correlations 
persist between the lipids within the bilayer.  This is a consequence of the 
important property that momentum is conserved for the SELM dynamics and 
that momentum can be transferred between lipids through the hydrodynamic 
fields.  This is in contrast to the Langevin dynamics where momentum is 
simply ceded locally to a fixed ambient medium.  As a consequence of the 
conservation of momentum in the SELM dynamics, the drag is expected to 
result in locally induced hydrodynamic flows.  To characterize more
precisely this behavior, we consider the pair correlation tensor 
for the motion of two lipids diffusing within the bilayer.

\subsection{Spatial Correlations between Lipids within a Vesicle Bilayer}

\begin{figure}[h]
\centering
  \includegraphics[width=8cm]{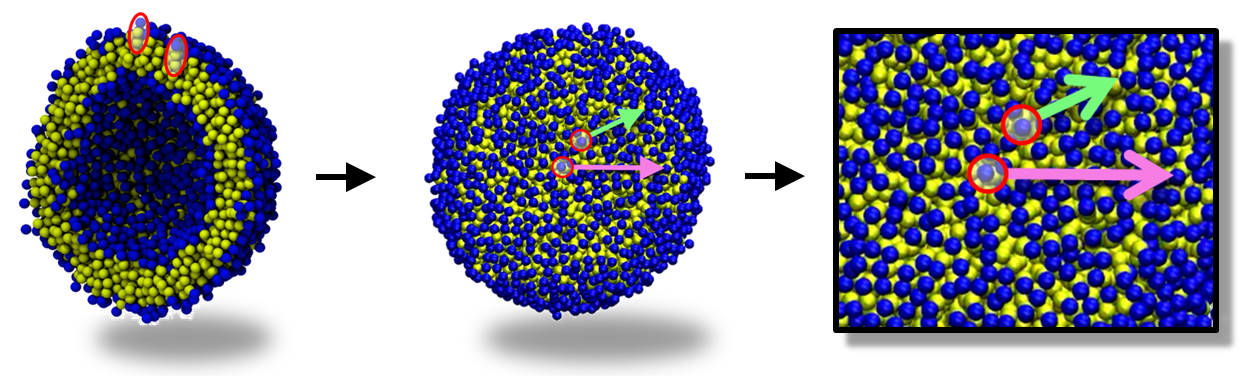}
  \caption{Analysis of Spatial Correlations between Lipids within a Vesicle Bilayer}  
\label{fig_pair_corr_schematic}
\end{figure}

\begin{figure}[h]
\centering
  \includegraphics[width=7cm]{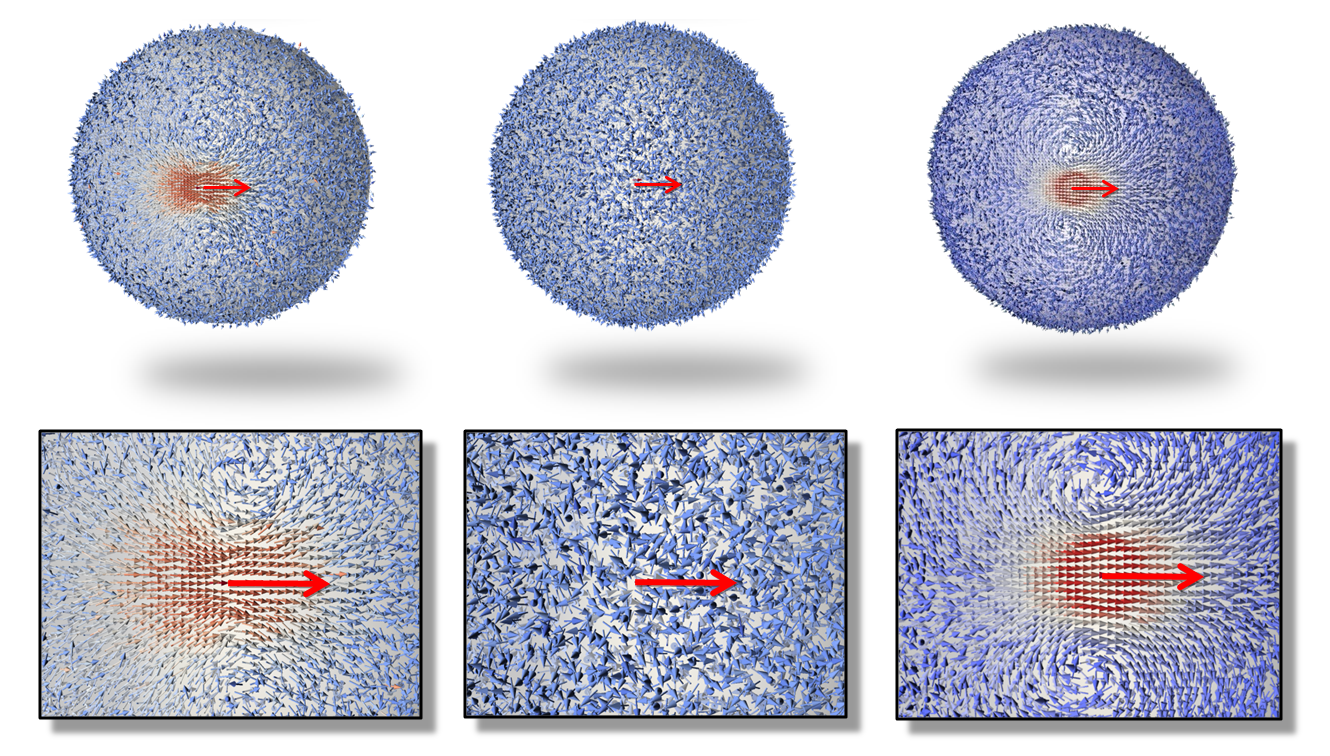}
  \caption{Spatial Correlations between Lipids within a Vesicle Bilayer.  Shown
are results for Langevin with small drag $\Upsilon=0.06$ $m_0/\tau$ (left),
Langevin with Stokes drag $\Upsilon = 7210$ $m_0/\tau$ (middle), 
and SELM Fluctuating Hydrodynamics with $\Upsilon=7210$ $m_0/\tau$ (right).
For the Langevin dynamics with small drag and the SELM Fluctuating Hydrodynamics,
lipid flows are observed exhibiting correlations 
resembling vortex-like structures.
}  
\label{fig_pair_corr_results}
\end{figure}

\begin{figure}[h]
\centering
  \includegraphics[width=7cm]{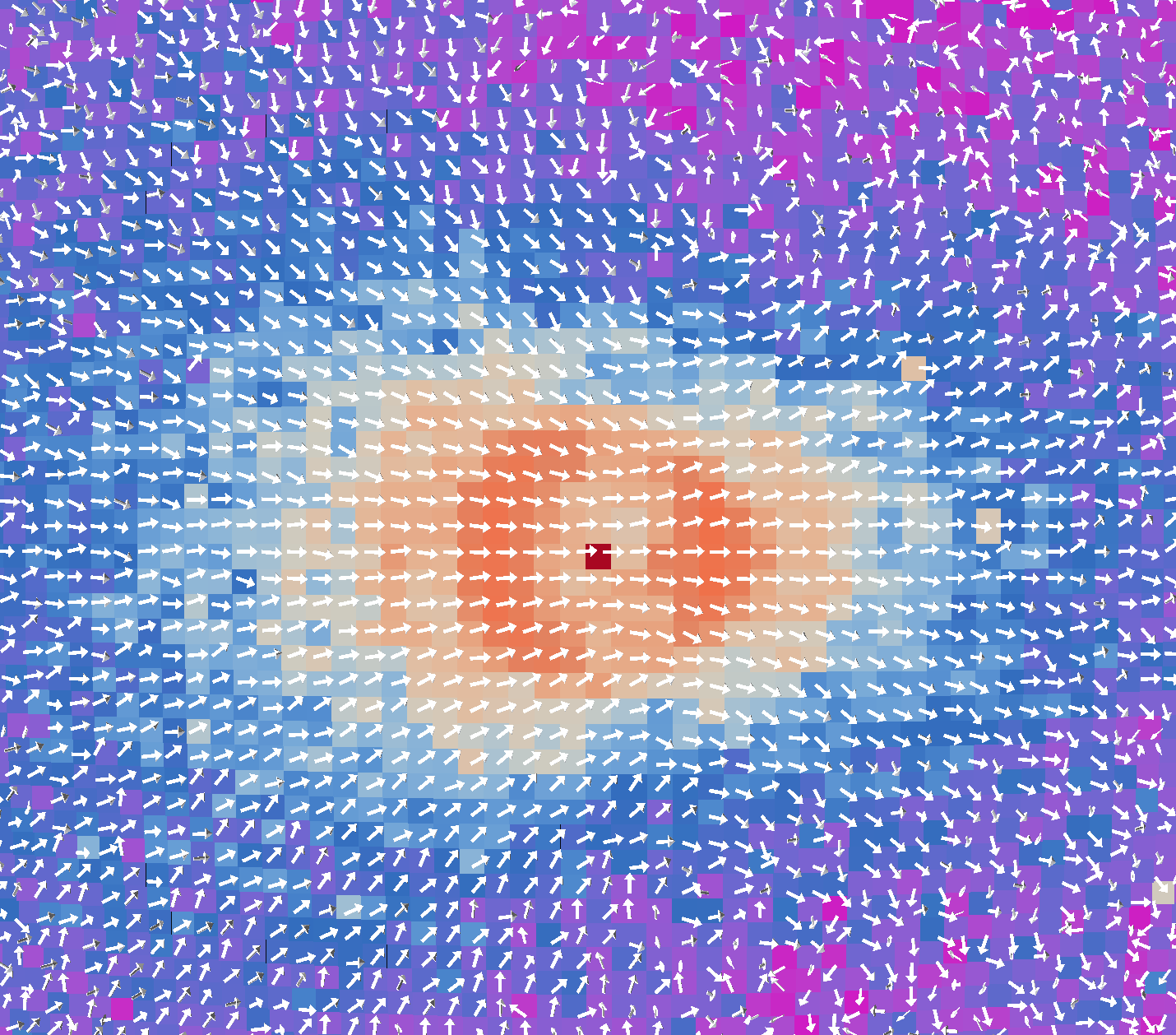}
  \caption{Lipid Flow Correlations for Langevin with Small Drag $\Upsilon = 0.06$ $m_0/\tau$.}  
\label{fig_vortex_Langevin_15_results}
\end{figure}

\begin{figure}[h]
\centering
  \includegraphics[width=7cm]{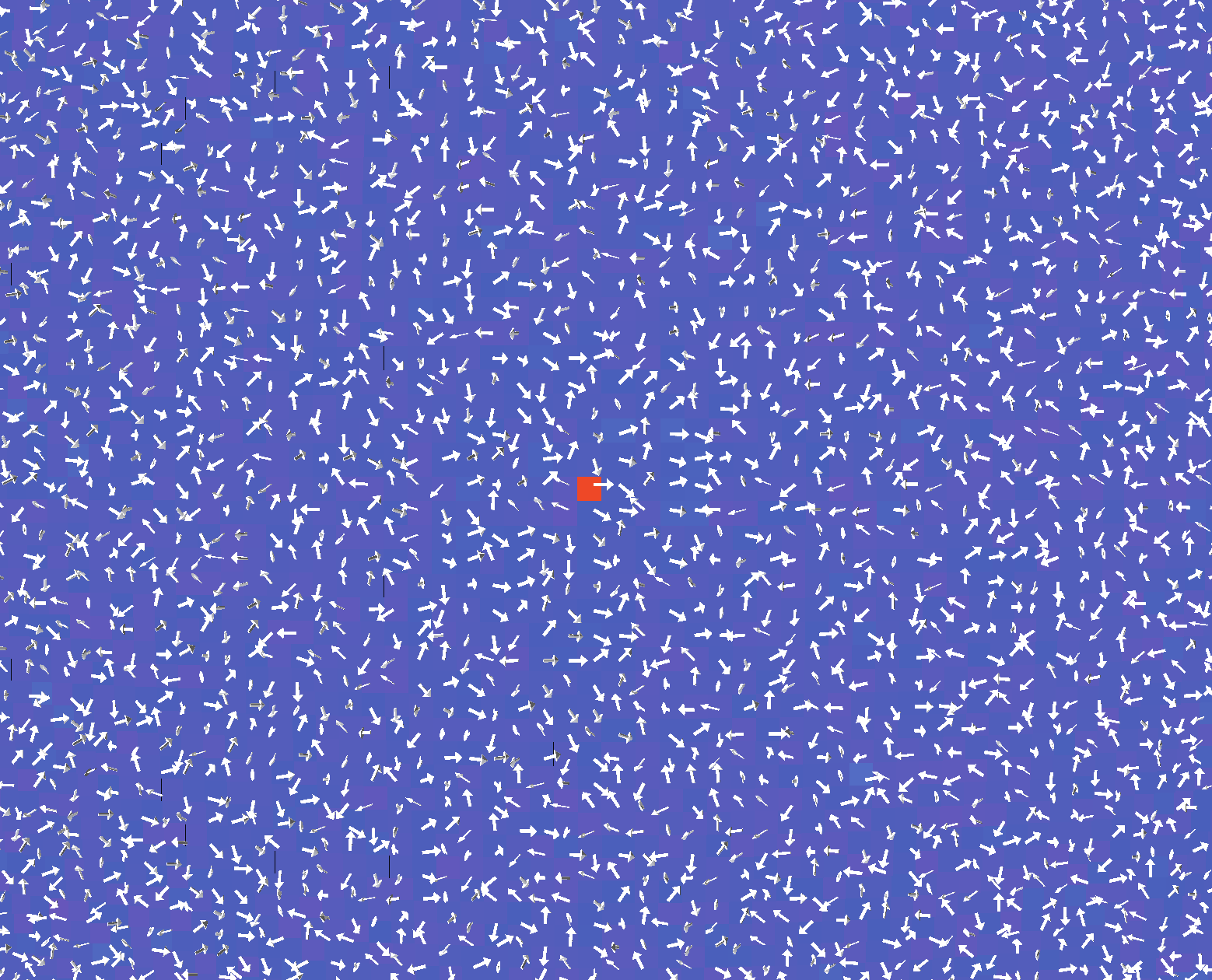}
  \caption{Lipid Flow Correlations for Langevin with Stokes Drag $\Upsilon = 7210$ $m_0/\tau$.}  
\label{fig_vortex_Langevin_Stokes_results}
\end{figure}

\begin{figure}[h]
\centering
  \includegraphics[width=7cm]{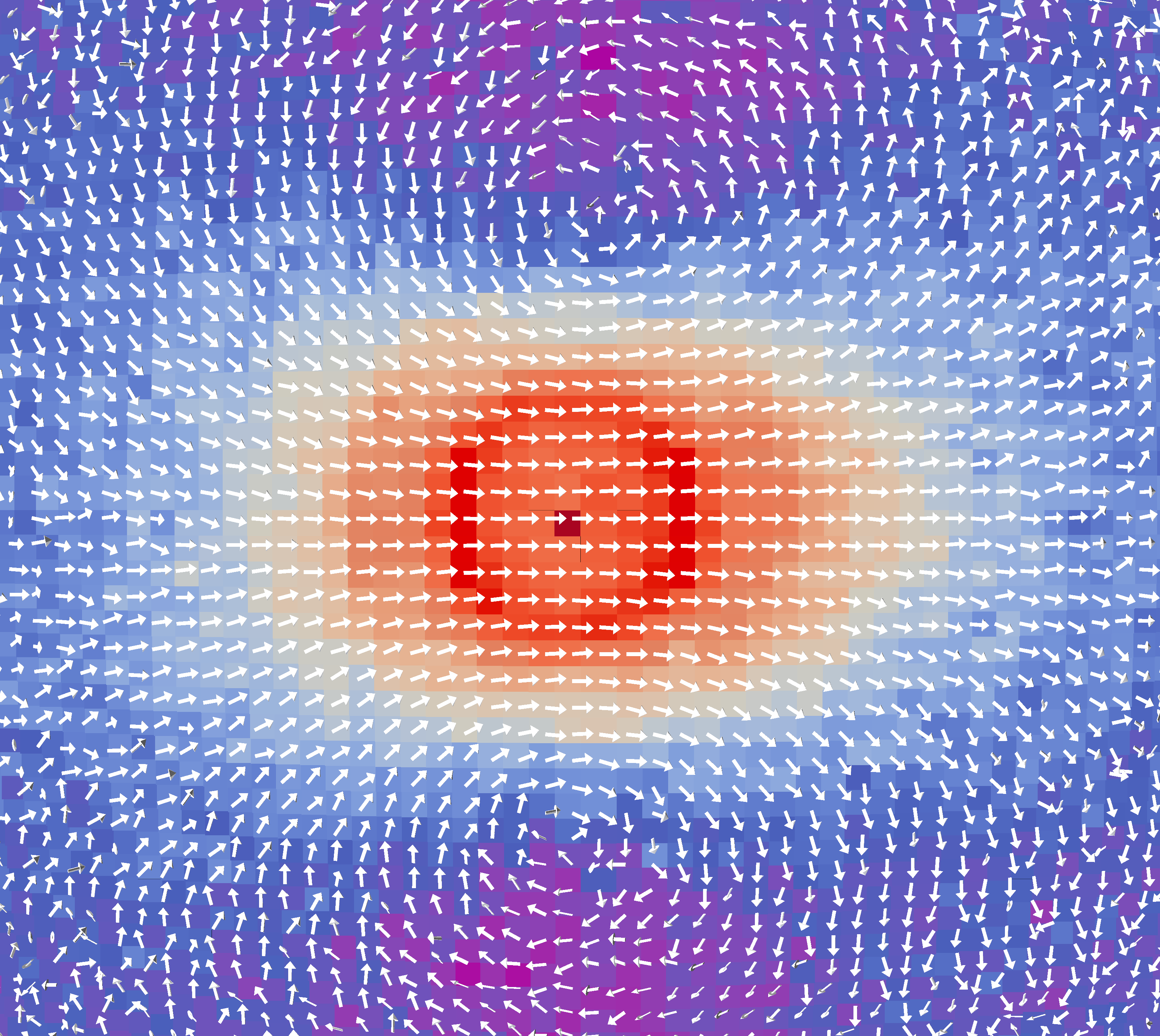}
  \caption{Lipid Flow Correlations for SELM Fluctuating Hydrodynamics $\Upsilon = 7210$ $m_0/\tau$.}  
\label{fig_vortex_SELM_results}
\end{figure}

To characterize more precisely the spatial correlations between the motions of 
lipids diffusing within the bilayer, we consider the pair correlation tensor
defined by
$\Psi(\mb{r}) = \left\langle\Delta_{\mb{r}}{X}\Delta_0{X}^T\right\rangle$.
In this notation, the displacements $\Delta{X}$ are taken over the time 
$\delta{t}$ and the subscript specifies the displacement vector from 
the center-of-mass of a reference lipid to the center-of-mass of a 
second lipid within the bilayer, see Figure~\ref{fig_pair_corr_schematic}.
The vector field $\mb{w} = \Psi \mb{e}_1$ provides a characterization of
the correlations in the flow of lipids within the bilayer.  These spatial 
correlations are shown for the two types of Langevin dynamics considered 
and the SELM dynamics,
see Figure~\ref{fig_pair_corr_results}.
We find that for the Langevin dynamics with small drag coefficient,
there is a significant lipid flow structure exhibited within 
the bilayer, see left of Figure~\ref{fig_pair_corr_results} and~\ref{fig_vortex_Langevin_15_results}.  
This appears to be a consequence of the interial dynamics
and collision events that occur to transfer momentum laterally within the 
bilayer.  In contrast, the Langevin dynamics with Stokes drag greatly 
surpresses this momentum transfer between lipids and there are no discernible 
spatial correlations, see middle of Figure~\ref{fig_pair_corr_results} and~\ref{fig_vortex_Langevin_Stokes_results}.    
In the SELM dynamics with Stokes drag, interesting lipid flows are 
exhibited having a vortex-like structure, see right of 
Figure~\ref{fig_pair_corr_results} and~\ref{fig_vortex_SELM_results}.  
In fact, very similar vortex-like flow structures have been observed in 
explicitly solvated simulations of lipid bilayers, see~\citep{Falck2008,Apajalahti2010}.
Interestingly, similar lipid flows and spatial correlations have been offered as an 
explanation for recent neutron scattering experiments~\cite{Armstrong2011}. 
These results indicate that to obtain realistic lipid dynamics within bilayers it is 
important to conserve momentum and to incorporate the solvent mediated momentum 
transfer between lipids.  

\clearpage

\section{Conclusion}
\label{Sec_Conclusion}

We have introduced a fluctuating hydrodynamics approach for 
dynamical studies of implicit-solvent coarse-grained lipid models.
We have presented a general framework and specific methods 
for how to couple coarse-grained degrees of freedom with stochastic hydrodynamic
fields.  We have shown that our fluctuating hydrodynamics lipid model yields 
bilayer dynamics that differ markedly from conventional Langevin 
dynamics.  We have shown that the conservation of momentum 
and the hydrodynamic transfer of momentum between the lipids  
plays an important role in producing coherent flows of lipids 
within the bilayer.  In contrast, the Langevin dynamics with a 
comparable drag greatly suppresses momentum transfer 
between the lipids and results in only short-range correlations.
For diffusion within vesicles, we have shown that our fluctuating 
hydrodynamics lipid model has interesting spatial correlations 
that exhibit a vortex-like flow structure.  The results show the 
promise of the methods to capture important hydrodynamic
mediated effects previously observed in explicit solvent 
molecular dynamics simulations and in recent neutron scattering
experiments~\cite{Armstrong2011,Falck2008,Apajalahti2010}. 
We expect the SELM fluctuating hydrodynamics methods to provide 
powerful new approaches for performing investigations of 
dynamical processes in lipid bilayers utilizing implicit-solvent 
coarse-grained models.

\section{Acknowledgements}
The author P.J.A. acknowledges support from research grant NSF CAREER - 0956210.  We also acknowledge support 
from the W. M. Keck Foundation to Y.W. and support from the UCSB Center for Scientific Computing 
NSF MRSEC (DMR-1121053) and UCSB MRL NSF CNS-0960316.  We would also like to thank the Kavli Institute for 
Theoretical Physics under NSF PHY05-51164 for support to participate in the workshop benefiting this
work ``Physical Principles of Multiscale Modeling, Analysis and Simulation in Soft Condensed Matter.''
For stimulating discussions and useful suggestions concerning this work, we would also like to 
thank T. Chou, M. Deserno, S. Marrink, P. Pincus, and G. Huber. 

\bibliographystyle{siam}
\bibliography{paperBibliography}

\vspace{3cm}

\appendix

\end{document}